\documentclass[prl,aps,twocolumn,showpacs]{revtex4}
\usepackage{graphicx}
\usepackage{subfigure}
\begin{document}

\title{Sample-specific and Ensemble-averaged Magnetoconductance\\ of Individual Single-Wall Carbon Nanotubes}
\author{H.T.~Man}
\author{A.F.~Morpurgo}
\affiliation{Kavli Institute of NanoScience Delft, Faculty of
Applied Sciences, \\Delft University of Technology, Lorentzweg 1,
2628 CJ Delft, The Netherlands}
\date{\today}

\begin{abstract}
We discuss magnetotransport measurements on individual single-wall
carbon nanotubes (SWNTs) with low contact resistance, performed as a
function of temperature and gate voltage. We find that the
application of a magnetic field perpendicular to the tube axis
results in a large magnetoconductance of the order of $e^2/h$ at low
temperature. We demonstrate that this magnetoconductance consists of
a sample-specific and of an ensemble-averaged contribution, both of
which decrease with increasing temperature. The observed behavior
resembles very closely the behavior of more conventional
multi-channel mesoscopic wires, exhibiting universal conductance
fluctuations and weak localization. A theoretical analysis of our
experiments will enable to reach a deeper understanding of
phase-coherent one-dimensional electronic motion in SWNTs

\end{abstract}


\maketitle

The study of magnetotransport is a powerful tool for the
investigation of the electronic properties of mesoscopic systems
\cite{GenMeso}. In single-wall carbon nanotubes (SWNTs), however,
only limited magnetotransport studies have been performed until
now. These studies have mainly focused on effects originating from
the coupling of the magnetic field to the electron spin
\cite{Spin_Cobden1,Spin_Cobden2,Spin_Liang,Kondo_Nygard}, whereas
effects originating from coupling to the orbital motion have not
received much attention until very recently
\cite{Orb_McEuen,AB_Dai}. This is probably because the lateral
dimensions of SWNTs are very small, which may have led to the idea
that orbital effects could play a relevant role only in the
presence of an extremely high magnetic field, beyond experimental
reach \cite{MRTheory_Ando}.

Many basic questions regarding the magnetoconductance of SWNTs
have so far been left unanswered by the lack of systematic
experimental investigations. For instance, it is not known what
kind of magnetoconductive response would be observed even in
simple experimental configurations, such as a metallic SWNTs in
the presence of magnetic field applied normally to the tube axis.
More specifically, it is not known what determines the magnetic
field scale required to induce an appreciable change in an
individual SWNT conductance, how large this change would be, and
what is the dominant microscopic mechanism responsible for its
occurrence. These questions are particularly relevant because
SWNTs are considered to be truly one-dimensional conductors in
which electron-electron interactions are expected to play a key
role \cite{LLTheory_Kane}, with considerable experimental evidence
for Luttinger liquid behavior having been reported during the past
few years \cite{LL_Postma}.

\begin{figure}[b]
  \centering
  \includegraphics[width=8.5cm]{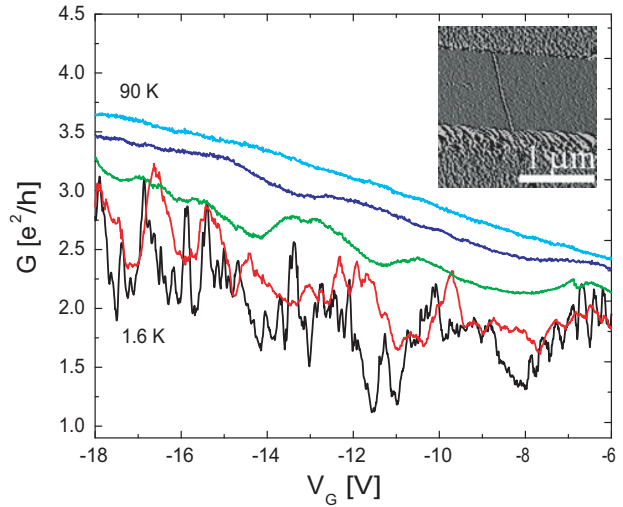}
  \caption{Gate voltage dependence of the conductance of an individual
  SWNT (shown in the inset) measured at different temperatures
  (1.6 K - bottom curve -, 4.2 K, 20 K, 45 K, and 90 K - top curve).
  Note the aperiodic conductance fluctuations whose amplitude
  decreases with increasing temperature.}\label{Results_Gate}
\end{figure}

In this letter we report systematic experimental investigations of
magnetotransport through individual SWNTs, which demonstrate the
presence of a large magnetoconductance of orbital origin, appearing
already on a rather small field scale. All our experiments were
performed on metallic SWNTs with low contact resistance, to prevent
the occurrence of Coulomb-blockade at low temperature. In these
samples we observe large ($\simeq e^2/h$) aperiodic
magnetoconductance fluctuations upon the application of a magnetic
field perpendicular to the tube axis, whose shape strongly depends
on the gate voltage. By averaging over different magnetoconductance
curves measured on the same SWNT at different values of the gate
voltage, we show that it is possible to suppress the aperiodic
fluctuations and to reveal the presence of a positive
magnetoconductance. We further show that the magnitude of the
"ensemble-averaged" magnetoconductance and of the aperiodic
fluctuations decreases with increasing temperature. The qualitative
behavior of the observed magnetoconductance is interpreted in terms
of orbital coupling of the magnetic field, which affects the quantum
interference of electron waves in the SWNTs, resulting in
magneto-induced conductance fluctuations and in the suppression of
weak localization. Our results call for a theoretical analysis of
these phenomena, whose occurrence in SWNTs had not been anticipated,
which will enable the quantitative determination of the yet unknown
phase-coherence time in SWNTs and will deepen our understanding of
the interplay between quantum interference and electron-electron
interaction in these systems.

\begin{figure}[t]
  \centering
  \includegraphics[width=8.5cm]{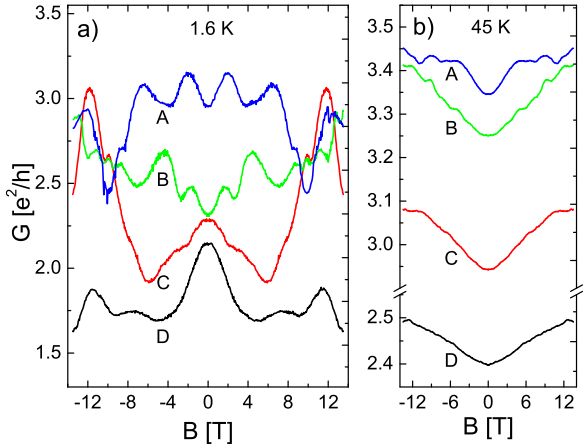}\\
  \caption{a) Individual magnetoconductance traces measured at 1.6 K,
  for different values of the gate voltage (A, $V_\mathrm{G}$ = -16.8 V; B,
  $V_\mathrm{G}$ = -15 V; C, $V_\mathrm{G}$ = -12 V; D, $V_\mathrm{G}$ = -6.8 V), showing large
  aperiodic fluctuations, whose precise shape strongly depends on $V_\mathrm{G}$.
  Note that both a maximum and a minimum is observed around $B=0$. b)
  Magnetoconductance traces measured at $T =$ 45 K, for the same values
  of $V_\mathrm{G}$: at higher temperature, the magnetoconductance has always a
  minimum around $B=0$.}\label{Results_Magn1K}
\end{figure}

The devices used in our investigations consist of an individual
metallic SWNT in between two contacts, prepared on a degenerately
doped silicon substrate (also acting as gate electrode) coated with
a 500 nm-thick thermally grown oxide layer. The SWNTs are deposited
by means of a chemical vapor deposition process
\cite{NTgrowth_Kong}; the electrodes consist of a Pd layer (10 nm
thick) covered with a second Au$_{0.6}$Pd$_{0.4}$ layer (15 nm
thick), resulting in a high yield of samples having a contact
resistance lower than the quantum resistance \cite{PdContact_Dai}.
The overall details of the fabrication process are very similar to
those described in Ref.~\cite{Fab_Morpurgo}. Magnetotransport
measurements were performed on different devices which show
identical qualitative behavior. Here we will discuss data obtained
from one single sample, representative of this behavior.

Figure \ref{Results_Gate} shows the conductance $G$ of an individual
SWNT measured as a function of gate voltage $V_\mathrm{G}$, at
different temperatures. With varying $V_\mathrm{G}$, the conductance
exhibits pronounced aperiodic oscillations whose peak-to-peak
amplitude is $\delta G \simeq e^2/h$ at 1.6 K and decreases with
increasing temperature. This behavior is commonly observed when
measuring low-temperature transport through SWNTs, with highly
transparent contacts. The aperiodic oscillations are attributed to
phase-coherent electron waves interfering randomly in the presence
of disorder \cite{DirtyFabry_Kong}. Whereas transport through
disorder-free SWNTs with transparent contacts has been investigated
in some detail (e.g., analysis of a Fabry-Perot interference pattern
\cite{Fabry_Liang}), no systematic study addressing the influence of
disorder has been reported to date.

Figure \ref{Results_Magn1K}(a) shows the magnetoconductance of the
same individual SWNT for which the $G(V_\mathrm{G})$ curves are
plotted in Fig.~\ref{Results_Gate}. The data were taken at 1.6 K
with the magnetic field applied perpendicular to the tube axis, for
different values of gate voltage. It is apparent that for all values
of gate voltage the magnetic field induces large changes in the
conductance, the details of which are strongly dependent on the
value of $V_\mathrm{G}$. The change in $V_\mathrm{G}$ needed to
substantially change the measured magnetoconductance (MC) curve is
approximately 0.3 V, which corresponds well to the correlation
voltage of the $G(V_\mathrm{G})$ curve measured at $B=0$
(Fig.~\ref{Results_Gate}).

For all values of gate voltage, the MC curves are symmetric, as
expected for measurements performed in a two-terminal configuration
in the presence of time-reversal symmetry \cite{GenMeso} and, at 1.6
K, either a maximum or a minimum in the MC is observed around $B=0$
depending on the value of $V_\mathrm{G}$. The peak-to-peak amplitude
of the magnetic field-induced conductance fluctuations is $\delta G
\simeq 0.5 \ e^2/h$, comparable to the fluctuations induced by
changes in the gate voltage. The characteristic magnetic field scale
of the aperiodic fluctuations is $\delta B \simeq 3-4$ T.

\begin{figure}[b]
  \centering
  \includegraphics[width=8.5cm]{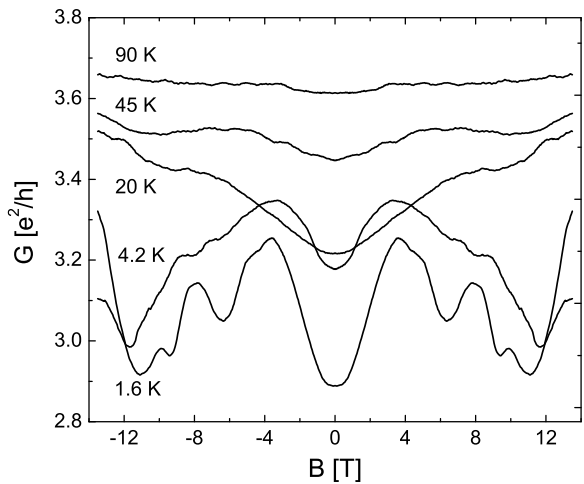}\\
  \caption{Magnetoconductance curves measured at $V_\mathrm{G}$ = -17.7 V for
  different temperatures, showing that the magnitude of the aperiodic conductance
  fluctuations decreases with increasing temperature.}\label{Results_OscTemp}
\end{figure}

As the temperature is increased, the magnitude of the magnetic
field-induced conductance fluctuations decreases, similarly to what
happens to the gate voltage-induced fluctuations (see
Fig.~\ref{Results_OscTemp}). In addition, at higher temperature ($T
> 40$ K) the MC exhibits a qualitatively different behavior.
Specifically, at sufficiently low field the MC is positive for all
values of gate voltage, so that a conductance minimum is always
observed around $B=0$ T as is shown in Fig.~\ref{Results_Magn1K}(b).

The observed conductance fluctuations, which are induced by a
change in gate voltage or in magnetic field and have an amplitude
close to $\delta G \simeq e^2/h$, resemble very closely universal
conductance fluctuations (UCF) seen in multi-channel mesoscopic
wires \cite{GenMeso}. In these wires, the low-temperature MC
originates from the magnetic flux piercing the system, which
affects the phase-coherent propagation of electrons in a
disordered medium through random Aharonov-Bohm phases acquired by
the electronic waves. The characteristic magnetic field scale for
these fluctuations is given by $\delta B \approx\Phi_0/S$, where
$\Phi_0 = h/e $ and $S$ is the area of the samples where the
phase-coherent propagation of electronic waves takes place. In our
sample, this condition corresponds to a value of $\delta B \approx
\Phi_0/Ld \simeq 3$ T, which is in good agreement with our
experimental findings (here $L \simeq 1\ \mu$m and $d = 1.5$ nm
are the tube length and diameter respectively).

An interpretation of the MC fluctuations in terms of random
interference of electronic waves, similar to UCF, also provides a
natural explanation for the temperature dependence observed in our
experiments. The magnitude of the aperiodic fluctuation decreases
with increasing temperature because of two mechanisms. First, the
increase in thermal energy results in the population of a broader
window of states around the Fermi energy. Second, at sufficiently
high temperature the phase-coherence length $l_\phi$ becomes smaller
than the length of the SWNT, so that the tube behaves as a
collection of $L/l_\phi$ independent phase-coherent segments. Both
these mechanisms result in the averaging of the random conductance
oscillations, causing their suppression.

In conventional mesoscopic conductors, UCF are always accompanied by
weak localization and it is interesting to see whether the effect of
weak localization on the MC of SWNTs can also be observed
experimentally. In a fully phase-coherent conductor, UCF and weak
localization have comparable magnitude, which makes it difficult to
separate the two phenomena in a single MC measurement. However, upon
ensemble averaging, random MC fluctuations can be suppressed whereas
weak localization (WL) remains unaffected \cite{UCFTheory_Lee1}.

For this reason, we have performed an ensemble average of the MC
of an individual SWNT using as ensemble $N=34$ MC traces measured
at different values of $V_\mathrm{G}$ in the interval -6 V to -18
V. This procedure is motivated by the analogy with conventional
mesoscopic wires, for which it has been theoretically proven that
a sufficiently large change in Fermi energy (induced in our case
by the gate voltage) is equivalent to a complete change in
impurity configuration, in so far as the conductance oscillations
are concerned \cite{UCFTheory_Lee2}. In the experiment, the
separation in $V_\mathrm{G}$ between adjacent MC traces is $0.3 $
V, corresponding approximately to the position of the half-width
in the autocorrelation function of $G(V_\mathrm{G})$.

\begin{figure}[t]
  \centering
  \includegraphics[width=8.5cm]{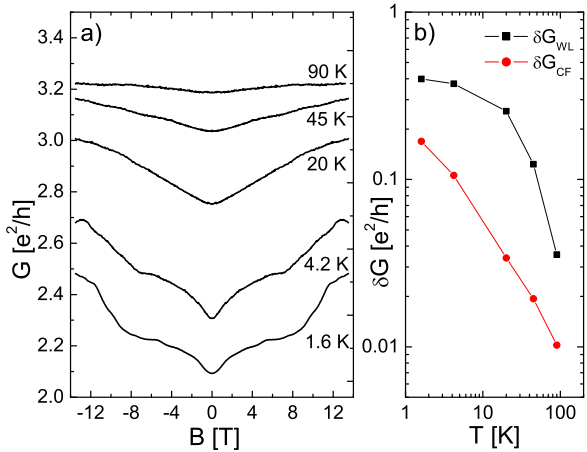}\\
  \caption{a) Ensemble-averaged magnetoconductance curves measured
  at different temperatures. In all cases the average is taken over
  34 magnetoconductance traces measured at different values of $V_\mathrm{G}$.
  It is apparent that a positive magnetoconductance remains after
  averaging. b) amplitude of the aperiodic conductance fluctuations
  and of the weak-localization signal as a function of temperature.
  The lines are a guide to the eye.}\label{Results_EnsAv}
\end{figure}

The result of the ensemble average is shown in
Fig.~\ref{Results_EnsAv}(a). It is apparent that the
ensemble-averaged MC is always positive, as it is expected if WL is
mechanism for the presence of MC. The magnitude of this positive MC
decreases gradually with increasing temperature, which is consistent
with the phase-coherence length in the nanotube becoming shorter
-also as expected- with increasing $T$. That there is essentially no
difference between the ensemble-averaged MC measured at 1.6 K and
4.2 K, indicates that in this temperature range $l_{\phi}$ is longer
than the length of the SWNT. Note also that at 1.6 K and 4.2 K some
structure is still visible superimposed on the background of the
positive MC, because the averaging procedure only suppresses the
magnitude of sample-specific fluctuations by a factor of $\sqrt{N}$
(approximately equal to 6 in our case). This structure is less
pronounced at higher temperature, when thermal effects and a
phase-coherence length shorter than the tube length also contribute
to averaging remnant sample-specific features.

Having understood the phenomena responsible for the observed MC, we
try to quantify their magnitude more accurately. The magnitude
$\delta G_\mathrm{{CF}}$ of the aperiodic conductance fluctuations
(CF) is quantified in terms of their rms value. This, we directly
calculate from the individual $G(B)$ traces after subtracting the
averaged magnetoconductance curve $\langle G(B) \rangle$, to remove
the effect of weak localization. Fig.~\ref{Results_EnsAv}b shows the
resulting temperature dependence of $\delta G_\mathrm{{CF}}$ (note
that, at each temperature, we have also averaged the values of
$\delta G_\mathrm{{CF}}$ obtained at different values of
$V_\mathrm{G}$ to improve the statistical accuracy of our result).
An estimate of the amplitude of weak-localization correction to the
conductance $\delta G_\mathrm{{WL}}$ is less straightforward at this
stage. Here, we simply take $\delta G_\mathrm{{WL}} = \langle G (14\
T) \rangle - \langle G (0\ T)\rangle$ (also plotted in
Fig.~\ref{Results_EnsAv}(b) as a function of temperature). However,
we emphasize that this value of $\delta G_\mathrm{{WL}}$ is not
accurate, both because 14 T (the largest field reachable in our
measurement set-up) may not be sufficient to fully suppress weak
localization and because remnant structure due to the aperiodic
fluctuations is still present in the average MC curve. More accurate
measurements of $\delta G_\mathrm{{WL}}$ require the ability to
apply a higher magnetic field and the use of a larger number of MC
curves in the calculation of the ensemble-averaged
magnetoconductance.

These results give a fully quantitative answer to the questions
that have been posed earlier. A large orbital magnetoconductance
is visible in metallic SWNTs already at low magnetic field, due to
the quantum interference of electron waves. The magnetoconductance
is large even though the electronic motion is nearly ballistic, as
indicated by the average transmission per channel which reaches
values as high as $\simeq 0.9$ (in Fig.~1 the conductance reaches
values larger than $3.5 \ e^2/h$). This shows that imperfections
in the SWNTs that cause only a minor amount of backscattering can
affect electronic interference strongly. Specifically, even small
imperfections can prevent the observation of a regular Fabry-Perot
interference pattern and result in a random pattern of conductance
oscillations (both as a function of gate voltage and of magnetic
field).

In spite of the fact that the observed phenomenology is very similar
to that of mesoscopic conductors of higher dimensionality
\cite{GenWL}, there exist no theoretical description of UCF and WL
in SWNTs enabling a quantitative analysis of our results. Note that
in this regard that the situation in SWNTs differs from that of
multi-walled nanotubes (MWNTs), where UCF and WL can be described
satisfactorily in terms of the conventional theory for quasi-1D
diffusive conductors \cite{UCFMWNT_Schonenberger}. For MWNTs this
works probably because the mean free path is comparable or shorter
than the tube circumference, which justifies the hypothesis of
diffusive motion. In addition, the geometrical dimensions of MWNTs
are considerably larger than the Fermi wavelength, which is a
necessary condition for the validity of the conventional theory
based on a semi-classical approximation.

Both these conditions are not fulfilled in SWNTs, and the
theoretical analysis of WL and UCF will have to be based on a
different starting point. We believe that in SWNTs the presence of
two bands at the Fermi energy is particularly important to account
for the magnetoresistance observed here. This is because with two
degenerate sub-bands even a small magnetic field can induce sub-band
mixing and change the scattering matrix (and thus the conductance)
of a SWNT sample. This mixing would of course not be possible if
SWNTs would have only one conduction channel (as one may naively
assume, since SWNTs are often referred to as an "ideal" realization
of 1D conductor), in which case we expect that UCF and WL would be
absent. Finally, theory will have to consider that in SWNTs
electron-electron interaction plays an important role, and may
require a Luttinger liquid description of the electron system. This
makes the analysis of WL and UCF in SWNTs particularly interesting,
since so far the theoretical analysis of these phenomena has been
confined to the case of Fermi liquids. In this regard, we also
emphasize that the analysis of our data in terms of a fully
quantitative theory should enable the determination of the
phase-coherence time and its temperature dependence in SWNTs , which
is currently unknown.

We thank T.M.~Klapwijk, C.~Strunk and Yu.~Nazarov for useful
discussions, and J.~Kong and C.~Dekker for help with the nanotube
growth. This work is supported by the "Stichting voor Fundamenteel
Onderzoek der Materie" (FOM). The work of A.F. Morpurgo is part of
the NWO Vernieuwingsimpuls 2000 program.


\begin{thebibliography}{99}
\bibitem{GenMeso} S. Datta, \textit{Electronic Transport in Mesoscopic Systems}, (Cambridge University Press, Cambridge, 2001).

\bibitem{Spin_Cobden1} D.H. Cobden \textit{et al.}, Phys. Rev.
Lett. \textbf{81}, 681 (1998).

\bibitem{Spin_Cobden2} D.H. Cobden and J. Nyg\aa rd, Phys. Rev.
Lett. \textbf{89}, 046803 (2002).

\bibitem{Spin_Liang} W.J. Liang, M. Bockrath, and H. Park, Phys. Rev.
Lett. \textbf{88}, 126801 (2002).

\bibitem{Kondo_Nygard} J. Nyg\aa rd, D.H. Cobden, and P.E.
Lindelof, Nature \textbf{408}, 342 (2000).

\bibitem{Orb_McEuen} E.D. Minot \textit{et al.}, Nature
\textbf{428}, 536 (2004).

\bibitem{AB_Dai} J. Cao, Q. Wang, M. Rolandi, and H.J. Dai, Phys. Rev. Lett. \textbf{93}, 216803 (2004).

\bibitem{MRTheory_Ando} T. Ando, J. Phys. Soc. Jpn. \textbf{73},
1273 (2004).

\bibitem{LLTheory_Kane} R. Egger and A.O. Gogolin, Phys. Rev. Lett.
\textbf{79}, 5082 (1997); C. Kane, L. Balents, and M.P.A. Fisher,
\textit{ibid}. \textbf{79}, 5086 (1997); H. Yoshioka  and A.A.
Odintsov, \textit{ibid}. \textbf{82}, 374 (1999);

\bibitem{LL_Postma} M. Bockrath \textit{et al.}, Nature
\textbf{397}, 598 (1999); Z. Yao, H.W.Ch. Postma, L. Balents, and C.
Dekker, \textit{ibid}. \textbf{402}, 273 (1999).

\bibitem{NTgrowth_Kong} J. Kong \textit{et al.}, Nature
\textbf{395}, 878 (1998).

\bibitem{PdContact_Dai} D. Mann \textit{et al.}, Nano Lett.
\textbf{3} (11), 1541 (2003); A. Javey \textit{et al.}, Nature
\textbf{424}, 654 (2003).

\bibitem{Fab_Morpurgo} H.T. Soh \textit{et al.}, Appl. Phys. Lett. \textbf{75},
627 (1999).

\bibitem{DirtyFabry_Kong} J. Kong \textit{et al.}, Phys. Rev.
Lett. \textbf{87}, 106801 (2001).

\bibitem{Fabry_Liang} W. Liang \textit{et al.}, Nature
\textbf{411}, 665 (2001).

\bibitem{UCFTheory_Lee1} P.A. Lee and A.D. Stone, Phys. Rev.
Lett. \textbf{55}, 1622 (1985).

\bibitem{UCFTheory_Lee2} P.A. Lee, A.D. Stone, and H. Fukuyama,
Phys. Rev. B \textbf{35}, 1039 (1987).

\bibitem{GenWL} See e.g. \textit{Mesoscopic
Phenomena in Solids}, eds. B.L. Altshuler, P.A. Lee, and R.A.
Webb, (North-Holland, New York, 1991) and references therein.

\bibitem{UCFMWNT_Schonenberger} C. Sch\"onenberger \textit{et al.},
Appl. Phys. A \textbf{69}, 283 (1999); B. Stojetz \textit{et al.},
New Journal of Physics \textbf{6}, 27 (2004).

\end{thebibliography}
\end{document}